\documentclass[a4paper,11pt]{article}
\usepackage{graphicx}
\newcommand{\be}{\begin{equation}}\newcommand{\ee}{\end{equation}}
\newcommand{\bea}{\begin{eqnarray}}\newcommand{\eea}{\end{eqnarray}}
\newcommand{\brr}{\begin{array}}\newcommand{\err}{\end{array}}
\newcommand{\ben}{\begin{enumerate}}\newcommand{\een}{\end{enumerate}}
\newcommand{\bib}{\bibitem}
\newcommand{\ba}{\begin{array}}
\newcommand{\ea}{\end{array}}

\def\lab{\label}\def\lan{\langle}
\def\lf{\left}

\def\non{\nonumber}\def\ran{\rangle}

\def\ri{\right}\def\wti{\widetilde}
\def\al{\alpha}\def\bt{\beta}
\def\De{\Delta}\def\ep{\epsilon}
\def\te{\theta}
\def\si{\sigma}
\def\om{\omega}
\newcommand{\mlab}[1]{\label{#1}}

\def\mass{{_{1,2}}}
\def\1{{_{1}}}\def\2{{_{2}}}
\def\bk{{\bf {k}}}

\begin{document}
\bibliographystyle{plain}

\begin{flushleft}
{\Large Observables in the Quantum Field Theory of neutrino mixing
and oscillations} \vspace{1cm}
\\
M.~Blasone$^{1}$, P.~Jizba$^{2}$, G.~Vitiello$^3$
\\
1) Institute f\"ur Theoretische Physik, Freie Universit\"at
Berlin, D-14195 Berlin, Germany \\
2) Institute of Theoretical Physics, University of Tsukuba,
Ibaraki 305-8571, Japan \\
3) Dipartimento di Fisica, INFM and INFN, Universit\`{a} di
Salerno, I-84100 Salerno, Italy
\end{flushleft}
\vspace{0.5cm}

\begin{abstract}
We report about recent results on the Quantum Field Theory of neutrino mixing and
oscillations. A discussion of the relevant observables
for flavor fields is given, leading
to oscillation formulas which exhibit corrections with respect to the usual ones.
\end{abstract}

\section{Introduction}

Recent experimental results \cite{experiments} have finally confirmed
the reality of
neutrino mixing and oscillations \cite{Pont}, after a long search.
Despite these successes, many theoretical aspects of this problem are still
unclear. In particular, difficulties arise already when attempting
to find a proper mathematical setting for the description of
mixing in the framework of  Quantum Field Theory (QFT).

This is indeed an important task since it
is well known \cite{Barg} that mixing of states with
different masses is not even allowed  in non-relativistic Quantum Mechanics (QM).
Nevertheless, the quantum mechanical treatment is the one usually
adopted for its simplicity and elegance \cite{Pont}.
A review of the problems connected
with the QM treatment of mixing and oscillations can be found in
Ref.\cite{Zralek}. Difficulties in the construction of the Hilbert
space for mixed neutrinos were pointed out in Ref.\cite{kimbook}.

Only recently \cite{BV95}-\cite{dispersion}
a consistent treatment of mixing and
oscillations\footnote{For a
discussion of the existing approaches to neutrino
oscillations in QFT, see Ref.\cite{Beuthe}.} in
QFT has been achieved and we report here on some of these developments.

\section{Neutrino mixing in Quantum Field Theory}

The quantization of mixed Dirac fields has been studied in detail in
Refs.\cite{BV95,BHV99,fujii,hannabuss,3flavors}. Here we report
the main results for the case
of two flavors.
Let us consider the usual mixing relations connecting the flavor
fields $\nu_{e}$ and $\nu_{\mu}$  with the free fields
$\nu_{1}$ and $\nu_{2}$  with definite masses $m_{1}$ and
$m_{2}$:
\bea\non
\nu_{e}(x) &=& \cos\te \; \nu_{1}(x)    \, +\, \sin\te \;\nu_{2}(x)
\\[2mm] \label{fermix}
\nu_{\mu}(x) &=&-   \sin\te\;\nu_{1}(x)  \, +\,
\cos\te\;\nu_{2}(x)\,, \eea
where $\te$ is the mixing angle. We can write Eqs.(\ref{fermix}) as
\bea\label{fermix2}
&&\nu_\si(x)\equiv G^{-1}_{\bf \te}(t)
\, \nu_j(x)\, G_{\bf \te}(t), \eea
with $(\si,j)=(e,1), (\mu,2)$ and where
\be\label{generator}
G_{\te}(t) = \exp\lf[\te \int d^{3}{\bf x}
\lf(\nu_{1}^{\dag}(x) \nu_{2}(x) - \nu_{2}^{\dag}(x) \nu_{1}(x)
\ri)\ri]\;. \ee
The generator of  mixing transformations has been studied
in Ref.\cite{BV95} where it was shown that its action on the vacuum $|0\ran_{1,2}$ for the
fields $\nu_{j}$ results in a new state $|0\ran_{e,\mu}$ -- the flavor vacuum:
\bea\label{flavac}
|0(t)\ran_{e,\mu}\,\equiv\,G_{\te}^{-1}(t)\;|0\ran_{1,2} \;,
\eea
which is orthogonal to $|0\ran_{1,2}$ in the infinite volume limit. In the following,
we will use
$|0\ran_{e,\mu}\equiv|0(0)\ran_{e,\mu}$.
The free fields  $\nu_j$ (j=1,2) can be quantized in the usual
way (we use $t\equiv x_0$) \cite{Um1}:
\bea\lab{nui}
&&{}\hspace{1cm}
\nu_{j}(x) = \sum_{r=1,2} \int
\frac{d^3\bk}{(2\pi)^\frac{3}{2}} \,
\lf[u^{r}_{{\bf k},j}(t) \al^{r}_{{\bf
k},j}\:+ v^{r}_{-{\bf k},j}(t) \bt^{r\dag }_{-{\bf k},j} \ri]\,
e^{i {\bf k}\cdot {\bf x}},
\qquad j=1,2 \;,
\eea
with $u^{r}_{{\bf k},j}(t)=e^{-i\om_{k,j} t}u^{r}_{{\bf k},j}$,
$v^{r}_{{\bf k},j}(t)=e^{i\om_{k,j} t}v^{r}_{{\bf k},j}$ and
$\om_{k,j}=\sqrt{{\bf k}^2+m_j^2}$.  The
anticommutation relations are the usual ones; the wave function
orthonormality and completeness relations are those of
Ref.\cite{BV95}.

By use of $G_{\bf \te}(t)$, the flavor fields can be expanded as:
\bea \label{flavfields}
&&\nu_\si(x)\,=\, \sum_{r=1,2} \int \frac{d^3
\bk}{(2\pi)^\frac{3}{2}} \lf[ u^{r}_{{\bf k},j}(t) \al^{r}_{{\bf k},\si}(t)
+    v^{r}_{-{\bf k},j}(t) \bt^{r\dag}_{-{\bf k},\si}(t) \ri]
e^{i {\bf k}\cdot{\bf x}}\,, \eea
with $(\si,j)=(e,1) , (\mu,2)$.
The flavor annihilation operators are defined as $\al^{r}_{{\bf
k},\si}(t) \equiv G^{-1}_{\bf \te}(t)\al^{r}_{{\bf k},j} G_{\bf
\te}(t)$ etc. They clearly act as annihilators for the flavor
vacuum Eq.(\ref{flavac}). For further use, it is helpful to
list them explicitly (see
also Ref.\cite{BV95}). In the reference frame with ${\bf k}=(0,0,|{\bf
k}|)$ the spins factorize and we have the simple expressions:
\bea \label{annih1}
&&{}\hspace{-1cm}
\al^{r}_{{\bf k},e}(t)=\cos\te\;\al^{r}_{{\bf
k},1}\;+\;\sin\te\;\lf( U_{{\bf k}}^{*}(t)\; \al^{r}_{{\bf
k},2}\;+\;\ep^{r}\;
V_{{\bf k}}(t)\; \bt^{r\dag}_{-{\bf k},2}\ri)\, ,
\\  \label{annih2}
&&{}\hspace{-1cm}
\al^{r}_{{\bf k},\mu}(t)=\cos\te\;\al^{r}_{{\bf
k},2}\;-\;\sin\te\;\lf( U_{{\bf k}}(t)\; \al^{r}_{{\bf
k},1}\;-\;\ep^{r}\;
V_{{\bf k}}(t)\; \bt^{r\dag}_{-{\bf k},1}\ri)\, ,
\\  \label{annih3}
&&{}\hspace{-1.2cm}
\bt^{r}_{-{\bf k},e}(t)=\cos\te\;\bt^{r}_{-{\bf
k},1}\;+\;\sin\te\;\lf( U_{{\bf k}}^{*}(t)\; \bt^{r}_{-{\bf
k},2}\;-\;\ep^{r}\;
V_{{\bf k}}(t)\; \al^{r\dag}_{{\bf k},2}\ri)\, ,
\\  \label{annih4}
&&{}\hspace{-1.2cm}\bt^{r}_{-{\bf k},\mu}(t)=\cos\te\;\bt^{r}_{-{\bf
k},2}\;-\;\sin\te\;\lf( U_{{\bf k}}(t)\; \bt^{r}_{-{\bf
k},1}\;+\;\ep^{r}\; V_{{\bf k}}(t)\; \al^{r\dag}_{{\bf k},1}\ri)\, ,
\eea
where $\ep^{r}=(-1)^{r}$ and
\bea &&
U_{{\bf k}}(t)\equiv u^{r\dag}_{{\bf k},2}(t)u^{r}_{{\bf
k},1}(t)=
v^{r\dag}_{-{\bf k},1}(t)v^{r}_{-{\bf k},2}(t)\,=\,
|U_{{\bf k}}|\;e^{i(\om_{k,2}-\om_{k,1})t}\, ,
\\ [2mm]
\mlab{2.37}
&&V_{{\bf k}}(t)\equiv \ep^{r}\;u^{r\dag}_{{\bf
k},1}(t)v^{r}_{-{\bf k},2}(t)= -\ep^{r}\;u^{r\dag}_{{\bf
k},2}(t)v^{r}_{-{\bf k},1}(t) \,=\,|V_{{\bf
k}}|\;e^{i(\om_{k,2}+\om_{k,1})t}\,,
\\ [2mm]\mlab{2.38}
&&{}\hspace{-1cm}
|U_{{\bf k}}|=\frac{|{\bf k}|^{2} +(\om_{k,1}+m_{1})(\om_{k,2}+m_{2})}{2
\sqrt{\om_{k,1}\om_{k,2}(\om_{k,1}+m_{1})(\om_{k,2}+m_{2})}}
\quad;\quad
|V_{{\bf k}}|=\frac{ (\om_{k,1}+m_{1}) - (\om_{k,2}+m_{2})}{2
\sqrt{\om_{k,1}\om_{k,2}(\om_{k,1}+m_{1})(\om_{k,2}+m_{2})}}\, |{\bf k}|\,.
\eea
In Eqs.(\ref{annih1})-(\ref{annih4}) a rotation is combined with a
Bogoliubov transformation, where the Bogoliubov
coefficients satisfy $|U_{{\bf k}}|^{2}+|V_{{\bf k}}|^{2}=1$.
Similar results hold for Majorana
and boson fields  \cite{bosonmix,binger,neutral}.

\section{Observables for mixed neutrinos}

In the previous Section we have seen how to quantize mixed fermion
(neutrino) fields, leading
to the expansion (\ref{flavfields}) for the fields with definite
flavor. We have also seen
that the vacuum structure is affected by the action of the mixing
generator (\ref{generator}) which
results in the flavor vacuum $|0\ran_{e,\mu}$ and in the non-trivial structure of the
flavor annihilation/creation operators (\ref{annih1})-(\ref{annih4}).

The question now is to see what are the physical implications of these
mathematical structures. To this end we address the question of what are
the observable quantities
for mixed fields.
Let us start with a discussion of the flavor states \cite{BHV99,BPT02,neutral}.
By definition, these states have definite flavor charge
and so, for neutrino and antineutrino states (denoted
as $|\nu_\si\ran$ and $|{\bar \nu_\si}\ran$) we should
have\footnote{In the following, we work in the Heisenberg picture.}
\bea\label{eigenvect1}
Q_\si \, |\nu_\si\ran \,= \, |\nu_\si\ran \quad;\qquad
Q_\si\,|{\bar \nu_\si}\ran \,= \, - |{\bar \nu_\si}\ran\,.
\eea
The flavor charges for mixed fields have been studied in detail in
Refs.\cite{currents,bosonmix,3flavors}.
In the present case of mixing of two Dirac fields, we obtain:
\bea
Q_\si(t)  \,\equiv \, \int d^3 {\bf x}\,\nu_\si^\dag(x)\nu_\si(x) \,=\,
\sum_{r}\int d^3 {\bf k} \lf( \alpha^{r\dag}_{{\bf
k},\sigma}(t) \alpha^{r}_{{\bf k},\sigma}(t)\, -\,
\beta^{r\dag}_{-{\bf k},\sigma}(t)\beta^{r}_{-{\bf
k},\sigma}(t)\ri)\,,\qquad\sigma= e,\mu,
\eea
where $Q_e(t) \, + \,Q_\mu(t) \, = \, Q$ with $Q$ being the total (conserved)
$U(1)$ charge \cite{currents}.
Thus the flavor charges
are diagonal in the flavor ladder operators. This is evident when we
realize how they are related to the Noether charges\footnote{These are the
$U(1)$ charges separately
conserved  for the free fields $\nu_j$. We have: $Q_1+Q_2 = Q$.} $Q_j$ \cite{currents}:
\bea\label{chaconn}
&&{}\hspace{3cm}
Q_\si(t) \,=\, G_\te^{-1}(t)\,Q_j \, G_\te(t)
\,,\qquad \qquad(\si,j)=(e,1),(\mu,2).\eea

We thus are led to the following definition for a  neutrino state with definite flavor:
\be \label{eneutrino}
|\nu_\si\rangle \equiv
\alpha^{r \dagger}_{{\bf k},\si}(0)|0\rangle_{e,\mu} =
G_\theta^{-1}(0) \alpha^{r \dagger}_{{\bf k},j}|0\rangle_\mass\;.
\ee
with similar expressions for antineutrinos. Clearly, the state
(\ref{eneutrino}) satisfies the requirement of Eq.(\ref{eigenvect1}).
Moreover, we can see that $|\nu_\si\rangle $ also satisfies
\bea
{\bf P}_\sigma(0)\,|\nu_\si\rangle \,= \,{\bf k}\,|\nu_\si\rangle
\eea
where the momentum operator for the mixed fields is defined as
\cite{neutral} ($\sigma = e,\mu$):
\bea \non
{}\hspace{-.5cm}
{\bf P}_\sigma(t)& =&\int d^3{\bf x} \,\nu^\dagger_\sigma(x)\,
(-i\nabla)\,\nu_\sigma(x) \, =\,  G_\te^{-1}(t)\,P_j \, G_\te(t)
\\
& =& \int d^3 {\bf k} \sum_r
\frac {\bf k} {2} \left( \alpha^{r\dagger}_{{\bf k},\sigma}(t)
\alpha^{r}_{{\bf k},\sigma}(t)  -
\alpha^{r\dagger}_{{-\bf k},\sigma}(t)\alpha^{r}_{{-\bf k},\sigma}(t) +
\beta^{r\dagger}_{{\bf k},\sigma}(t)\beta^{r}_{{\bf k},\sigma}(t)
-\beta^{r\dagger}_{-{\bf k},\sigma}(t)\beta^{r}_{-{\bf k},\sigma}(t)
\right).
\eea

Note that the usually employed Pontecorvo states $|\nu_e\ran = \cos\theta \,|\nu_1\ran +
\sin\theta\,|\nu_2\ran$ and $|\nu_{\mu}\ran  = -
\sin\theta\,|\nu_1\ran + \cos\theta\,|\nu_2\ran $ are not eigenstates of the flavor charge
neither of the momentum operator (see also Ref.\cite{Giunti:2001kj})
and thus are not consistently defined within QFT.

\section{Oscillation formulas}

We now use the above results to derive oscillation formulas.
Let us consider the case of an
electron neutrino state $|\nu_e\rangle \equiv
\alpha^{r \dagger}_{{\bf k},e}|0\rangle_{e,\mu}$. At time $t\neq 0$,
this is not anymore eigenstate of the
flavor charge operators. We obtain
$\;_{e,\mu}\langle 0|Q_\si(t)| 0\rangle_{e,\mu}\; =\; 0$ and
\bea
\label{charge2}
&& {\cal Q}_{{\bf k},\si}(t) \,\equiv\,
\langle \nu_e|
Q_\sigma(t) |\nu_e\rangle
\,=\,\lf|\lf \{\al^{r}_{{\bf k},\si}(t), \al^{r
\dag}_{{\bf k},e}(0) \ri\}\ri|^{2} \;+ \;\lf|\lf\{\bt_{{-\bf
k},\si}^{r \dag}(t), \al^{r \dag}_{{\bf k},e}(0) \ri\}\ri|^{2}\,.
\eea
Charge conservation is obviously ensured at any time: ${\cal
Q}_{{\bf k},e}(t) + {\cal Q}_{{\bf k},\mu}(t)\; = \; 1$. The oscillation formulas for
the flavor charges are \cite{BHV99}
\bea\label{enumber}
{}\hspace{-1cm}
{\cal Q}_{{\bf k},e}(t)&=& 1 \,-\, \sin^{2}( 2 \theta)\, |U_{{\bf k}}|^{2} \; \sin^{2}
\lf( \frac{\omega_{k,2} - \omega_{k,1}}{2} t \ri)
+  \sin^{2}( 2 \theta)\, |V_{{\bf
k}}|^{2} \; \sin^{2} \lf( \frac{\omega_{k,2} + \omega_{k,1}}{2}
t \ri)  \, ,
\\[3mm] \label{munumber}
{}\hspace{-1cm}
{\cal Q}_{{\bf k},\mu}(t)&=&  \sin^{2}( 2 \theta)\, |U_{{\bf k}}|^{2} \; \sin^{2}
\lf( \frac{\omega_{k,2} - \omega_{k,1}}{2} t \ri)
+\sin^{2}( 2 \theta)\,
|V_{{\bf k}}|^{2} \; \sin^{2} \lf( \frac{\omega_{k,2} + \omega_{k,1}}{2}
t \ri) \, .
\end{eqnarray}
This result is exact. The difference with respect to
the Pontecorvo formula \cite{Pont} is in the energy dependent amplitudes
and in the additional oscillating terms.
The usual
QM formulas \cite{Pont},
are approximately recovered in the relativistic limit ($k\gg \sqrt{m_\1m_\2}$)
where we obtain (for $\te=\pi/4$):
\bea
&&{}\hspace{-2cm}
{\cal Q}_{{\bf k},\mu}(t)
\simeq    \, \lf( 1- \frac{(\De m)^2}{4 k^2}\ri)
\sin^2 \lf[ \frac{\De m^2}{4 k} \, t \ri]
\,+  \, \frac{(\De m)^2}{4 k^2} \sin^2 \lf[\lf(k +
\frac{m_\1^2+m_\2^2}{4k}\ri) \, t \ri]\,.
\eea
%

Similar results are obtained when we consider the expectation values of the
momentum operator at a time $t\neq 0$. We have indeed
${}_{e,\mu}\lan 0 | {\bf P}_\si(t) | 0\ran_{e,\mu}\, =\, 0$ and
\bea\label{momentum2}
&&\frac{\langle \nu_e|{\bf P}_\si(t)| \nu_e\ran}
{\langle \nu_e|{\bf P}_\si(0)| \nu_e\ran}
 \;=\, \lf|\lf \{\al^{r}_{{\bf k},\si}(t), \al^{r
\dag}_{{\bf k},e}(0) \ri\}\ri|^{2} \;+ \;\lf|\lf\{\bt_{{-\bf
k},\si}^{r \dag}(t), \al^{r \dag}_{{\bf k},e}(0) \ri\}\ri|^{2}
\,,
\eea
which is the same expression obtained for the charges Eq.(\ref{charge2}). Note that
the momentum operator is well defined for Majorana fields, whereas the flavor
charge operators vanish for neutral fields \cite{neutral}.

\section{Discussion and conclusions}

We have seen  how the fields $\nu_e$ and $\nu_\mu$ can be
expanded in the same spinor bases as $\nu_1$ and $\nu_2$,
viz. Eq.(\ref{flavfields}).
However, such a choice
is actually a special one and a more
general possibility exists \cite{fujii}.
Indeed, in  the expansion (\ref{flavfields}) one could
use eigenfunctions with arbitrary masses $\mu_\sigma$  and write
the flavor fields as \cite{fujii}:
\begin{eqnarray}\label{exnuf2}
{}\hspace{-1cm}
&& \nu_{\sigma}(x) \,=\, \sum_{r=1,2} \int \frac{d^3
\bk}{(2\pi)^\frac{3}{2}}  \lf[
u^{r}_{{\bf k},\sigma} {\widetilde \alpha}^{r}_{{\bf k},\sigma}(t)
+ v^{r}_{-{\bf k},\sigma} {\widetilde \beta}^{r\dag}_{-{\bf
k},\sigma}(t) \ri]  e^{i {\bf k}\cdot{\bf x}} ,
\end{eqnarray}
where $u_{\sigma}$ and $v_{\sigma}$ are the
eigenfunctions with mass $\mu_\sigma$. We denote by a tilde the
generalized flavor operators introduced in Ref.\cite{fujii}.
The expansion Eq.(\ref{exnuf2}) is more
general than the one in Eq.(\ref{flavfields}) since the latter
corresponds to the particular choice $\mu_e\equiv m_1$, $\mu_\mu
\equiv m_2$.
Thus the Hilbert space for the flavor fields
is not unique: an infinite number of vacua
can be generated by introducing
the arbitrary mass parameters $\mu_\si$. It is obvious
that physical quantities must not depend on these parameters.
Similar results are valid for bosons, see Ref.\cite{bosonmix}.
It can be explicitly checked that ($\sigma,\rho = e,\mu$) \cite{remarks}:
\begin{equation}\label{miracle}
 \lf|\lf \{{\wti \al}^{r}_{{\bf k},\si}(t), {\wti \al}^{r
\dag}_{{\bf k},\rho}(t') \ri\}\ri|^{2} \;+ \;\lf|\lf\{{\wti \bt}_{{-\bf
k},\si}^{r \dag}(t), {\wti \al}^{r \dag}_{{\bf k},\rho}(t') \ri\}\ri|^{2} \, =\,
\lf|\lf \{\al^{r}_{{\bf k},\si}(t), \al^{r
\dag}_{{\bf k},\rho}(t') \ri\}\ri|^{2} \;+ \;\lf|\lf\{\bt_{{-\bf
k},\si}^{r \dag}(t), \al^{r \dag}_{{\bf k},\rho}(t') \ri\}\ri|^{2}\, ,
\end{equation}
which ensures the cancellation of the arbitrary mass parameters in the
expectation values (\ref{charge2}),(\ref{momentum2}).
Thus the independence of expectation values on the arbitrary parameters provides
a criterion for the selection of the observables for mixed fields. Indeed,
 the number operators for mixed fields are not good observables
since their expectation values do depend on the arbitrary mass parameters.

In conclusion, we have discussed in this report how to properly
define observable quantities
for mixed (Dirac) fields in the context of Quantum Field Theory. We derived
oscillation
formulas which exhibit corrections with respect to the usual ones.

\section*{Acknowledgements}
We acknowledge  the ESF Program COSLAB,  INFN and INFM for
partial financial support.

\end{document}